\documentclass[12pt]{article}
\usepackage{epsfig,a4wide}
\usepackage{graphicx}
\usepackage[all,2cell,ps]{xy}

\newcommand{\dmu}{\partial_\mu}

\newcommand{\lc}{{\cal L}}

\newcommand{\jc}{{\cal J}}
\newcommand{\lm}{{\cal M}}

\newcommand{\oh}{{1\over2}}

\newcommand{\sqd}{\sqrt{2}}
\newcommand{\sqt}{\sqrt{3}}
\newcommand{\sqs}{\sqrt{6}}
\newcommand{\sqdt}{\sqrt{2 \over 3}}

\newcommand{\squt}{{1 \over \sqt}}
\newcommand{\squd}{{1 \over \sqd}}

\title{Axial Resonances in the Open and Hidden Charm Sectors}

\author{D.~Gamermann$^1$, E.~Oset$^1$ \\ 
{\small{\it $^1$Departamento de F\'{\i}sica Te\'orica and IFIC,
Centro Mixto Universidad de Valencia-CSIC,}}\\
{\small{\it Institutos de
Investigaci\'on de Paterna, Aptdo. 22085, 46071 Valencia, Spain}}
}

\begin{document}

\maketitle

\abstract{A $SU(4)$ flavor symmetrical Lagrangian is constructed for the interaction of the pseudo-scalar mesons with the vector mesons. $SU(4)$ symmetry is broken to $SU(3)$ by suppression of terms in the Lagrangian where the interaction should be driven by charmed mesons. Chiral symmetry can be restored by setting this new $SU(4)$ symmetry breaking parameters to zero. Unitarization in coupled channels leads to the dynamical generation of resonances. Many known axial resonances can be identified including the new controversial $X(3872)$ and the structure found recently by Belle around 3875 MeV in the hidden charm sector. Also new resonances are predicted, some of them with exotic quantum numbers.}

\section{Introduction}

 Recent years have been very exciting for the hadron spectroscopy because of the discovery of many new and controversial states that do not fit well the interpretation of baryons as $qqq$ states or mesons as $q\bar q$ states. In particular two charmed resonances discovered by BaBar [\ref{babar}] and confirmed by other experiments [\ref{cleo}], [\ref{belle1}], [\ref{belle2}], the $D_{s0}(2317)$ and $D_{s1}(2460)$ have animated the debate about non $q\bar q$ mesons. Also non-strange partners of these resonances have been observed [\ref{belle3}], [\ref{focus}].

 The predictions for the masses of these states with quark model potentials already existed [\ref{isgur}] and turned out to be off by more than 100 MeV. The fact that the $D_{s0}(2317)$ lies just below the $DK$ threshold and the $D_{s1}(2460)$ just below the $D^*K$ threshold made many theoreticians speculate that these states could be meson molecules [\ref{mol1}], [\ref{lutz1}], [\ref{lutz2}], [\ref{chiang1}], [\ref{chiang2}], [\ref{chiang3}]. Others support a tetraquark assignment [\ref{4q1}], [\ref{4q2}], [\ref{4q3}], or usual $q\bar q$ states with more sophisticated quark model potentials or within QCD sum rules calculations [\ref{qqbas1}], [\ref{qqbas2}], [\ref{qqbas3}], [\ref{qqbas4}] and there is also the possibility of admixture between these configurations [\ref{mixt1}], [\ref{mixt2}], [\ref{mixt3}].

 In the hidden-charm sector also new controversial resonances have been found. In particular the $X(3872)$, observed in four different experiments [\ref{expx1}], [\ref{expx2}], [\ref{expx3}], [\ref{expx4}], has attracted much attention. The narrow width of this state makes its interpretation as a usual charmonium $c\bar c$ state very difficult. For this resonance too, many exotic theoretical interpretations have been investigated such as tetraquarks, hybrids and molecules [\ref{theox1}], [\ref{theox2}], [\ref{theox3}], [\ref{theox4}], [\ref{theox5}]. For a good review on heavy mesons one can refer to [\ref{review}].

 In this work unitarization in coupled channels is used to explore the pseudo-scalar meson interaction with vector mesons. In the works of Kolomeitsev [\ref{lutz1}] and Guo [\ref{chiang1}] a similar approach has been done using a Lagrangian based on heavy quark chiral symmetry that allowed the investigation only of the open-charm sector and constrained the interaction for only light pseudo-scalars with heavy vector mesons. For our phenomenological model we construct a Lagrangian based on $SU(4)$ flavor symmetry and this symmetry is broken to $SU(3)$ by suppressing exchanges of heavy mesons in the implicit Weinberg-Tomozawa term. Chiral symmetry can be restored from our model by setting the $SU(4)$ symmetry breaking parameters to zero. This new Lagrangian, based on the ideas of a previous paper [\ref{gamer}], includes also the possibility to investigate the hidden-charm sector and the interaction of heavy pseudo-scalars with light vector mesons, which enriches the spectrum of axial resonances generated.

 The paper is organized as follows: in the next section the construction of the Lagrangian is explained in detail and also the mathematical framework for solving the scattering equations in a unitarized approach is presented. Section 3 is devoted to the presentation and discussion of the results and section 4 contemplates overview and conclusions.

\section{Mathematical Framework}

 We will start by constructing two fields, one for the $SU(4)$ 15-plet of pseudo-scalars and another one for the 15-plet of vector mesons:

\begin{eqnarray}
 \Phi&=&\sum_{i=1}^{15}{\varphi_i \over \sqd}\lambda_i \nonumber =\\ &=&\left( \begin{array}{cccc}
 {\pi^0 \over \sqd}+{\eta \over \sqs}+{\eta_c \over \sqrt{12}} & \pi^+ & K^+ & \bar D^0 \\ & & & \\
 \pi^- & {-\pi^0 \over \sqd}+{\eta \over \sqs}+{\eta_c \over \sqrt{12}} & K^0 & D^- \\& & & \\
 K^- & \bar K^0 & {-2\eta \over \sqs}+{\eta_c \over \sqrt{12}} & D_s^- \\& & & \\
 D^0 & D^+ & D_s^+ & {-3\eta_c \over \sqrt{12}} \\ \end{array} \right) \\
 \cal{V}_\mu&=&\sum_{i=1}^{15}{v_{\mu i} \over \sqd}\lambda_i \nonumber =\\ &=&\left( \begin{array}{cccc}
 {\rho_\mu^0 \over \sqd}+{\omega_\mu \over \sqs}+{J/\psi_{ \mu} \over \sqrt{12}} & \rho^+_\mu & K^{*+}_\mu & \bar D^{*0}_\mu \\ & & & \\
 \rho^{*-}_\mu & {-\rho^0_\mu \over \sqd}+{\omega_\mu \over \sqs}+{J/\psi_{\mu} \over \sqrt{12}} & K^{*0}_\mu & D^{*-}_\mu \\& & & \\
 K^{*-}_\mu & \bar K^{*0}_\mu & {-2\omega_\mu \over \sqs}+{J/\psi_{\mu} \over \sqrt{12}} & D_{s\mu}^{*-} \\& & & \\
 D^{*0}_\mu & D^{*+}_\mu & D_{s\mu}^{*+} & {-3J/\psi_\mu \over \sqrt{12}} \\ \end{array} \right). \label{vecfie}
\end{eqnarray}

 Now for each one of these fields a vector current is build:

\begin{eqnarray}
J_\mu&=&(\dmu \Phi)\Phi-\Phi\dmu\Phi \\
\cal{J}_\mu&=&(\dmu \cal{V}_\nu)\cal{V}^\nu-\cal{V}_\nu\dmu \cal{V}^\nu .\label{curj}
\end{eqnarray}

 The Lagrangian is then constructed by connecting the two currents:

\begin{eqnarray}
\lc={-1\over 4f^2}Tr\left(J_\mu\cal{J}^\mu\right). \label{lagini}
\end{eqnarray}

Note that the $\omega$ appearing in eq. (\ref{vecfie}) is not the physical $\omega$ but $\omega_8$. The addition of a singlet state, which is diagonal and proportional to the identity matrix in the representation of eq. (\ref{vecfie}) does not give any contribution to $\cal{J}_\mu$ in eq. (\ref{curj}) and hence does not modify the Lagrangian in eq. (\ref{lagini}). However, there is a caveat since the singlet and octet mix strongly to give the $\omega$ and $\phi$ states which have different masses. In the results section we shall come back to this problem and will discuss the effect of this mixing.

Next step is to break $SU(4)$ symmetry, this will be done by suppressing exchanges of heavy mesons. To identify these terms, one should first decompose each one of the fields into its $SU(3)$ components:

\begin{eqnarray}
 \Phi&=&\left( \begin{array}{cc}
 \phi_8+{1\over\sqrt{12}}\phi_1 \hat 1_3 & \phi_3 \\ \phi_{\bar3} & -{3\over\sqrt{12}}\phi_1
 \\ \end{array} \right) \\
 \cal{V}_\mu&=&\left( \begin{array}{cc}
 V_{8\mu}+{1\over\sqrt{12}}V_{1\mu} \hat 1_3 & V_{3\mu} \\ V_{\bar3 \mu} & -{3\over\sqrt{12}}V_{1\mu}
 \\ \end{array} \right) .
\end{eqnarray}
 The $\hat 1_3$ is the 3x3 identity matrix and the fields $\phi_i$ and $V_{i\mu}$ contain the meson fields for each $i$-plet of $SU(3)$ into which the 15-plet of $SU(4)$ decomposes:

\begin{eqnarray*}
 \phi_8&=&\left( \begin{array}{ccc}
 {\pi^0 \over \sqd}+{\eta \over \sqs} & \pi^+ & K^+  \\
 \pi^- & {-\pi^0 \over \sqd}+{\eta \over \sqs} & K^0 \\
 K^- & \bar K^0 & {-2\eta \over \sqs} \\
  \end{array} \right) \\
 \phi_3&=&\left(\begin{array}{c} \bar D^0 \\ D^-\\ D_s^- \end{array} \right) \\
\phi_{\bar 3}&=&\left(\begin{array}{ccc} D^0 & D^+ & D_s^+ \end{array} \right) \\
 \phi_1&=& \eta_c \\
 V_{8\mu}&=&\left( \begin{array}{ccc}
 {\rho^0_\mu \over \sqd}+{\omega_\mu \over \sqs} & \rho^+_\mu & K^{*+}_\mu  \\
 \rho^-_\mu & {-\rho^0_\mu \over \sqd}+{\omega_\mu \over \sqs} & K^{*0}_\mu \\
 K^{*-}_\mu & \bar K^{*0}_\mu & {-2\omega_\mu \over \sqs} \\
  \end{array} \right) \\
 V_{3\mu}&=&\left(\begin{array}{c} \bar D^{*0}_\mu \\ D^{*-}_\mu \\ D_{s\mu}^{*-} \end{array} \right) \\
V_{\bar 3 \mu}&=&\left(\begin{array}{ccc} D^{*0}_\mu & D^{*+}_\mu & D_{s\mu}^{*+} \end{array} \right) \\
 V_{1\mu}&=& J/\psi_{\mu}
\end{eqnarray*}

 In terms of the $SU(3)$ fields the Lagrangian reads:

\begin{eqnarray}
\lc&=&{-1\over 4f^2}\Big( Tr\bigg(J_{88\mu}\jc_{88}^\mu+J_{3\bar3\mu}\jc_{3\bar3}^\mu+J_{88_\mu}\jc_{3\bar3}^\mu+
 J_{3\bar3\mu}\jc_{88}^\mu+J_{83\mu}\jc_{\bar3 8}^\mu+\nonumber \\
 & &{2\over\sqrt{3}}(J_{83\mu}\jc_{\bar3 1}^\mu+ J_{13\mu}\jc_{\bar38}^\mu)+ 
 {4\over3}J_{13\mu}\jc_{\bar3 1}^\mu\bigg)+J_{\bar3 3\mu}\jc_{\bar3 3}^\mu+J_{\bar3 8\mu}\jc_{83}^\mu+\nonumber \\
 & & {2\over\sqrt{3}}(J_{\bar3 8\mu}\jc_{13}^\mu+J_{\bar3 1\mu}\jc_{83}^\mu)+
 {4\over3}J_{\bar3 1\mu}\jc_{13}^\mu\Big) . \label{lag}
\end{eqnarray}

 In eq.(\ref{lag}) the currents are defined as:

\begin{eqnarray}
J_{ij\mu}&=&(\dmu\phi_i)\phi_j-\phi_i\dmu\phi_j \label{curr1}\\
\jc_{ij\mu}&=&(\dmu V_{i\nu})V_j^\nu-V_{i\nu}\dmu V_j^\nu .\label{curr2}
\end{eqnarray}

The interaction in the Lagrangian of eq. (\ref{lag}) is usually visualized, in the vector-meson dominance picture, as a t-channel exchange of a vector meson in between the meson pairs. Since this Lagrangian is $SU(4)$ flavor symmetric, it assumes equal masses for the virtual vector meson exchanged in between the meson pairs. In order to break $SU(4)$ symmetry we will suppress the terms in this Lagrangian where the virtual vector meson exchanged is a heavy one. If the transferred momentum in between the meson pairs is neglected, the propagator of the exchanged meson is proportional to the inverse of its squared mass (${1\over m_V^2}$). Considering the terms where a light vector mesons is exchanged of the order of magnitude of the unit, the suppressed terms should be thought to be of the order $\left({m_L\over m_H}\right)^2$ where $m_L$ and $m_H$ are scales of the order of magnitude of the light and heavy vector mesons masses respectively.

Terms in the Lagrangian where the two currents of eqs. (\ref{curr1}) and (\ref{curr2}) have explicit charm quantum number should be suppressed since these currents can only be connected through the exchange of a charmed, and hence heavy, vector meson. There are still two terms in the Lagrangian where a hidden-charm meson can be exchanged in between the meson pairs. These are the terms involving only mesons belonging to the $SU(3)$ triplet and anti-triplet. In this case the interaction is driven simultaneously by light and heavy vector mesons and in order to correctly suppress these terms one should isolate each one of these contributions and suppress only the one coming from the heavy vector meson ($J/\psi$). These contributions have already been calculated in a previous paper [\ref{gamer}] and they can be either $1 \over 3$ and $2\over 3$ for light and heavy vector mesons respectively, if two equal currents are connected, or $-{1 \over 3}$ and $4\over 3$ for light and heavy vector mesons respectively, if two different currents are connected. So, following the steps in [\ref{gamer}], the corrected Lagrangian, accounting for the masses of the heavy vector mesons, reads as:

\begin{eqnarray}
\lc&=&{-1\over 4f^2}\Big( Tr\bigg(J_{88\mu}\jc_{88}^\mu+J_{3\bar3\mu}\jc_{3\bar3}^\mu+J_{88_\mu}\jc_{3\bar3}^\mu+
 J_{3\bar3\mu}\jc_{88}^\mu+\gamma J_{83\mu}\jc_{\bar3 8}^\mu+ \nonumber \\
& &{2\gamma\over\sqrt{3}}(J_{83\mu}\jc_{\bar3 1}^\mu+ J_{13\mu}\jc_{\bar38}^\mu)+
{4\gamma\over3}J_{13\mu}\jc_{\bar3 1}^\mu\bigg)+\psi J_{\bar3 3\mu}\jc_{\bar3 3}^\mu+ \nonumber \\
 & &\gamma J_{\bar3 8\mu}\jc_{83}^\mu+{2\gamma\over\sqrt{3}}(J_{\bar3 8\mu}\jc_{13}^\mu+J_{\bar3 1\mu}\jc_{83}^\mu)+
 {4\gamma\over3}J_{\bar3 1\mu}\jc_{13}^\mu \Big), \label{lag2}
\end{eqnarray}
 with $\gamma=\left(m_L\over m_H\right)^2$ and $\psi=-{1\over3}+{4\over3}\left(m_L\over m'_H\right)^2$.

The parameters $m_L$, $m_H$ and $m'_H$ should be chosen of the order of magnitude of a light vector-meson mass, charmed vector-meson mass and the $J/\psi$ mass, respectively. The masses of the light vector mesons vary in between 770 MeV, for the $\rho$ meson and 892 MeV for the $K^*$ mass, the charmed vector-mesons have masses 2008 MeV and 2112 MeV ($D^*$ and $D_s^*$, respectively) and the $J/\psi$ mass is approximately 3097 MeV. We will chose $m_L$=800 MeV, $m_H$=2050 MeV and $m'_H$=3 GeV. Changing these parameters over the whole allowed physical range (770 MeV $\leq$ $m_L$ $\leq$ 892 MeV and 2008 MeV $\leq$ $m_H$ $\leq$ 2112 MeV) has about 0.1\% effect over the pole position of the heavy resonances.

 Note that the first term of the Lagrangian in eq. (\ref{lag2}) is the same Lagrangian used in the works of Lutz [\ref{kolo}] and Roca [\ref{roca}] in the study of the low lying axial mesons and the third term is the same Lagrangian used in the works of Kolomeitsev [\ref{lutz1}] and Guo [\ref{chiang1}] when studying the open-charm sector. This matching sets the value ${1\over4f^2}$ as the coefficient in the Lagrangian of eq. (\ref{lagini}).

 From the Lagrangian in eq. (\ref{lag2}) one gets the transition amplitudes between an initial and a final state:

\begin{eqnarray}
\lm^C_{ij}(s,t,u)&=&{-\xi^C_{ij}\over4f^2}(s-u)\epsilon . \epsilon ' . \label{ampli}
\end{eqnarray}

 The super-index $C$ refers to the charge basis, and the labels $i$ and $j$ to the initial and final channels while $s$, $t$ and $u$ are the usual Mandelstam variables. In appendix A we give tables for the coefficients $\xi^I$ in an isospin basis.

 These amplitudes will first be projected in s-wave:

\begin{eqnarray}
V^I_{ij}(s)&=&\oh\int_{-1}^1 d(cos\theta) \lm^I_{ij}\bigg(s,t(s,cos\theta),u(s,cos\theta)\bigg)
\end{eqnarray}

 This potential will then be used as kernel in a Bethe-Salpeter equation, in an on-shell formalism [\ref{meme1}], [\ref{osetnd}], [\ref{osetkn}], [\ref{meisn}]. In this way the unitary T-matrix assumes the form [\ref{roca}]:

\begin{eqnarray}
 T&=&-({\hat 1} + V{\hat G})^{-1}V \overrightarrow{\epsilon}.\overrightarrow{\epsilon}' \label{bseq}
\end{eqnarray}
In this equation $\hat G$ is a diagonal matrix with each element given by:

\begin{eqnarray}
\hat G_l&=&G_l\left(1+{p^2\over3M^2}\right) \\
G_l&=&{1 \over 16\pi ^2}\biggr( \alpha _i+Log{m_l^2 \over \mu ^2}+{M_l^2-m_l^2+s\over 2s}
  Log{M_l^2 \over m_l^2}+\nonumber \\ 
 & &{p\over \sqrt{s}}\Big( Log{s-M_l^2+m_l^2+2p\sqrt{s} \over -s+M_l^2-m_l^2+
  2p\sqrt{s}}+Log{s+M_l^2-m_l^2+2p\sqrt{s} \over -s-M_l^2+m_l^2+  2p\sqrt{s}}\Big)\biggr)
  \label{loopf}
\end{eqnarray}
in the above equations $p$ is the three-momentum in the center of mass frame of the two mesons in channel $l$, while $M_l$ and $m_l$ are the masses of the vector and pseudo-scalar mesons respectively, $\alpha$ is the subtraction constant, which should be fitted as a free parameter and $\mu$ is a cut-off scale which we will set to 1.5 GeV. Note that $\alpha$ and $\mu$ are not independent, this justifies setting one to a fixed value and adjusting just the other one to data.

 The three-momentum $p$ is calculated with:

\begin{eqnarray}
p&=&{\sqrt{(s-(m_l+M_l)^2)(s-(m_l-M_l)^2)}\over 2\sqrt{s}} \label{trimom}
\end{eqnarray}

 Unitarity is ensured by the imaginary part of the loop function of equation (\ref{loopf}):

\begin{equation}
Im(G_l)=-{p\over 8\pi\sqrt{s}} \label{imloop}
\end{equation}

 When looking for poles in the complex plane one should be careful because of the cuts of the loop function 
 beyond each threshold. Bound states appear as poles over the real axis and below threshold in the
 first Riemann sheet. Resonances show themselves as poles above threshold and in the second
 Riemann sheet of the channels which are open.

 Over the real axis the discontinuity of the loop function is known to be two times its imaginary part [\ref{inoue}]
 so, knowing the value of the imaginary part of the loop function over the axis, eq. (\ref{imloop}), one can do a proper analytic continuation of it for the whole complex plane:

\begin{eqnarray}
G_{l}^{II}&=&G^{I}_{l}+i {p\over 4\pi\sqrt{s}}, \hspace{1cm} Im(p)>0
\end{eqnarray}
$G^{II}$ and $G^I$ refer to the loop function in the second and first Riemann sheets, respectively.

Until now our formalism worked only with stable particles, but in some cases one has a $\rho$ or a $K^*$ meson in the coupled channels, and these particles have relatively large widths. The consideration of the mass distributions of these particles can be relevant whenever thresholds are open thanks to this mass distribution.

In order to take this into account we follow the procedure of [\ref{sarkar}] and convolute the loop function with the spectral function of the particle, hence, using a new loop function:

\begin{eqnarray}
\tilde{G}(\sqrt{s},m,M_R)&=&{1\over N}\int_{(M_R-2\Gamma_R)^2}^{(M_R+2\Gamma_R)^2}d\tilde{M}^2\left({-1\over\pi}\right) Im{1\over \tilde{M}^2-M_R^2+iM_R\Gamma_R}.\nonumber \\
& & \hat{G}(\sqrt{s},m,\tilde{M}) \\
N&=&\int_{(M_R-2\Gamma_R)^2}^{(M_R+2\Gamma_R)^2}d\tilde{M}^2\left({-1\over\pi}\right) Im{1\over \tilde{M}^2-M_R^2+iM_R\Gamma_R}
\end{eqnarray}

In the next section we will comment further on this issue and present results, for the heavy resonances, by taking into account the finite width of the $\rho$ and $K^*$ vector mesons in the few cases where the generated resonances have important coupling to channels involving these mesons and a mass close to the threshold of these channels.

\section{Results}

 The 15-plet of $SU(4)$ breaks down into 4 multiplets of $SU(3)$:

\begin{equation}
15 \longrightarrow 1 \oplus 3 \oplus \bar 3 \oplus 8 .
\end{equation}

 Knowing this, one can study the $SU(3)$ structure of the interaction between pseudo-scalar and vector mesons. Table \ref{decompo} shows the $SU(3)$ decomposition of the interaction. The irreducible representations (irreps) marked with a $^*$ refer to the vector meson multiplet.

 \begin{table}
 \begin{center}
 \begin{tabular}{c||l}
 \hline
 charm & Interacting multiplets \\
 \hline 
 \hline
 & \\
 2 & $\bar 3 \otimes \bar 3^* \rightarrow 3 \oplus \bar 6$ \\
 \hline
 & \\
 1 & $\bar 3 \otimes 8^* \rightarrow \bar{15} \oplus \bar 3 \oplus 6 $ \\
   & $8 \otimes \bar 3^* \rightarrow \bar{15} \oplus \bar 3 \oplus 6 $ \\
   & $ \bar 3 \otimes 1^* \rightarrow \bar 3$ \\
   & $1 \otimes \bar 3^* \rightarrow \bar 3$ \\
 \hline
 & \\
 0 & $\bar 3 \otimes 3^* \rightarrow 8 \oplus 1$ \\
   & $ 3 \otimes \bar 3^* \rightarrow 8 \oplus 1$ \\
   & $1 \otimes 1^* \rightarrow 1$ \\
   & $8 \otimes 1^* \rightarrow 8$ \\
   & $1 \otimes 8^* \rightarrow 8$ \\
   & $8 \otimes 8^* \rightarrow 1 \oplus 8_s \oplus 8_a \oplus 10 \oplus \bar {10} \oplus 27$ \\
 \hline
 \end{tabular}
 \caption{$SU(3)$ decomposition of the interaction between pseudo-scalar and vector mesons in $SU(4)$. The sectors not shown in the table correspond to the $C=-1,-2$ states which are just charge
 conjugate states (antiparticles) from the ones shown.} \label{decompo}
 \end{center}
 \end{table}

 Since now one can differentiate between the vector and the pseudo-scalar representations in the irrep products, the coupled channel space gets enlarged with respect to the case of the scalar resonances and therefore, a much richer spectrum is generated for the axial resonances. In contrast with previous works of Kolomeitsev [\ref{lutz1}] and Guo [\ref{chiang1}] where light pseudo-scalar mesons are scattered off heavy vector mesons, in our work we have an enlarged coupled channel basis accounting also for channels where light vector mesons are scattered off heavy pseudo-scalars.

 Using $SU(3)$ isoscalar factors [\ref{gamer}], [\ref{su31}], [\ref{su32}] the $\xi^I_{ij}$ can be transformed to a $SU(3)$ basis, by means of which one knows in which multiplets there is attraction and, therefore, the possibility to generate resonances. In the C=2 sector the interaction is not attractive in any multiplet, in the C=1 sector there is attraction in the anti-triplets and in the sextets and in the C=0 sector the two octets and the singlet coming from $8 \otimes 8^*$ and the two heavy singlets from the $\bar 3 \otimes 3^*$ and the $3 \otimes \bar 3^*$ are attractive.

 Table \ref{channels} shows the channel content in each sector. We will work with states of defined charge conjugation or G-parity, where it applies.

 \begin{table}
 \begin{center}
 \begin{tabular}{c||c|c|l}
 \hline
Charm& Strangeness & I$^{G}(J^{PC})$& Channels \\
 \hline
 \hline
 & & & \\
 1 & 1 & $1(1^+)$ & $\pi D_s^*$, $D_s\rho$ \\
 & & &  $KD^*$, $DK^*$\\
\cline{3-4}
  & & & \\
  & & $0(1^+)$ & $DK^*$, $KD^*$, $\eta D_s^*$ \\
 & & & $D_s\omega$, $\eta_cD_s^*$, $D_sJ/\psi$ \\
\cline{2-4}
  & & & \\
  & 0 & $\oh(1^+)$ & $\pi D^*$, $D\rho$, $KD_s^*$, $D_s K^*$  \\
 & & & $\eta D^*$, $D\omega$, $\eta_c D^*$, $D J/\psi$\\
\cline{2-4}
  & & & \\
  & -1 & $0(1^+)$ & $DK^*$, $KD^*$ \\
 \hline
  & & & \\
 0 & 1 & $\oh(1^+)$ & $\pi K^*$, $K\rho$, $\eta K^*$, $K\omega$ \\
 & & & $\bar{D}D_s^*$, $D_s\bar{D^*}$, $KJ/\psi$, $\eta_c K^*$ \\
\cline{2-4}
 & & & \\
 & 0 & $1^+(1^{+-})$ & $\squd(\bar{K}K^*+c.c.)$, $\pi\omega$, $\eta \rho$\\
 & & & $\squd(\bar{D}D^*+c.c.)$, $\eta_c \rho$, $\pi J/\psi$ \\
\cline{3-4}
 & & & \\
 & & $1^-(1^{++})$ & $\pi\rho$, $\squd(\bar{K}K^*-c.c.)$, $\squd(\bar{D}D^*-c.c.)$\\
\cline{3-4}
 & & & \\
 & & $0^+(1^{++})$ & $\squd(\bar{K}K^*+c.c.)$, $\squd(\bar{D}D^*+c.c.)$, $\squd(\bar{D_s}D_s^*-c.c.)$ \\
\cline{3-4}
 & & & \\
 & & $0^-(1^{+-})$ & $\pi\rho$, $\eta\omega$, $\squd(\bar{D}D^*-c.c.)$, $\eta_c\omega$ \\
 & & & $\eta J/\psi$, $\squd(\bar{D_s}D_s^*+c.c.)$, $\squd(\bar{K}K^*-c.c.)$, $\eta_c J/\psi$ \\
\hline
 \end{tabular}
 \caption{Channel content in each sector} \label{channels}
 \end{center}
 \end{table}

 Our model assumes that isospin symmetry is exact so, all particles belonging to a same isospin multiplet have the same mass. For the pions we use $m_\pi$=138 MeV, for kaons $m_K$=495 MeV and for the eta $m_\eta$=548 MeV. In the heavy sector we use, for the pseudo-scalars $m_D$=1865 MeV, $m_{D_s}$=1968 MeV and $m_{\eta_c}$=2979 MeV. The masses of the light vector mesons are: $m_\rho$=771 MeV, $m_{K^*}$=892 MeV and $m_\omega$=782 MeV. While for the heavy vector mesons we use: $m_{D^*}$=2008 MeV, $m_{D_s^*}$=2112 MeV and $m_{J/\psi}$=3097 MeV.

 It is also possible to restore $SU(3)$ symmetry by setting the masses of all particles in a same $SU(3)$ multiplet to a common value. For this purpose we introduce the parameter $x$, $x=0$ is the case when $SU(3)$ symmetry is restored and $x=1$ the case we see in Nature with $SU(3)$ broken. The meson masses as a function of $x$ are given by:

\begin{eqnarray}
m(x)=\bar m+x(m_{phys.}-\bar m)
\end{eqnarray}
where $\bar m$ is the meson mass in the symmetric limit. We will use for the pseudo-scalars:

$\bar m_8$=430 MeV,
$\bar m_3$=1900 MeV and $\bar m_1$=1900 MeV, and for the vector mesons:

$\bar m_{8^*}$=800 MeV, $\bar m_{3^*}$=2050 MeV and $\bar m_{1^*}$=2050 MeV

 In our model we still have to fit the subtraction constants in the loop function. As done in our previous work [\ref{gamer}] we use two different values, $\alpha_L$ for channels involving just light mesons and $\alpha_H$ for channels involving at least one heavy meson. This is justified because the heavy and light sector nearly decouple from each other. The result of our fit is $\alpha_L=-0.8$ and $\alpha_H=-1.55$.

 We also use two different values for the meson decay constant $f$ appearing in the Lagrangian: for light mesons $f=f_\pi=93$ MeV, while for heavy mesons $f=f_D=165$ MeV [\ref{pdg}].

 When $SU(3)$ symmetry is restored it is possible to identify 3 poles and one cusp in the open charm sector. Two poles come from the two anti-triplets where the interaction was attractive, their positions are at 2432.63 MeV and (2535.07-i0.08) MeV. The other pole is broad and comes from one of the sextets at (2532.57-i199.36) MeV, the other sextet appears as a narrow cusp around 2700 MeV, it becomes a pole when the heavy light threshold at 2700 MeV moves because of the $SU(3)$ symmetry breaking.

\begin{figure}
\begin{center}
\includegraphics[angle=-0]{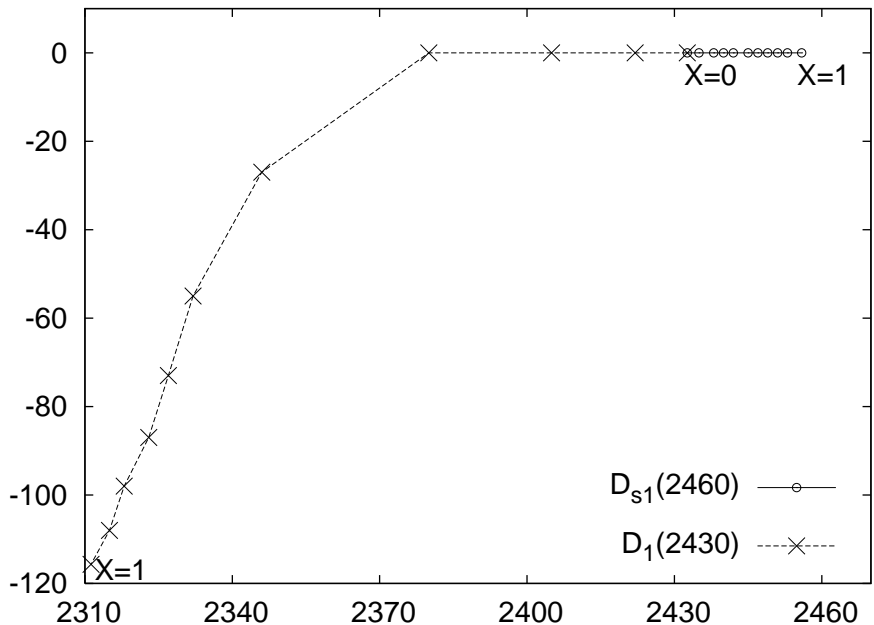}
\caption{Pole trajectories for one of the anti-triplets in the C=1 sector while breaking $SU(3)$ symmetry in steps of $\Delta x$=0.1. The two degenerate poles at $x$=0 become two different resonances at $x$=1, in the two extremes of the curve (real world).} \label{figure1}
\end{center}
\end{figure}

 Figure \ref{figure1} shows the pole trajectories for the anti-triplet starting at 2432.63 MeV while changing $x$ from 0 to 1 in steps of 0.1.

 In the hidden charm sector two octets and three singlets are expected, one light and two heavy. The two octets are nearly degenerated at 1161.06 MeV and 1161.37 MeV. In the work of Roca [\ref{roca}] the two octets are degenerated, but in our model the interaction with the heavy sector removes this degeneracy by a very small amount, indicating a weak coupling between the light and heavy sectors. The light singlet appears as a pole at 1055.77 MeV and the two heavy ones at 3867.59 MeV and (3864.62-i0.00) MeV this second one is not exactly a bound state as the others, but a narrow state with a width smaller than 1 KeV.

 Table \ref{resultsx1} shows the pole positions for the case $x=1$ within our model and the possible identification of each one.

\begin{table}
\begin{center}
\begin{tabular}{c|c||c|c|c|c|c}
\hline
C& Irrep &S&I$^G(J^{PC})$& RE($\sqrt{s}$) (MeV)& IM($\sqrt{s}$) (MeV)&Resonance ID\\
 & Mass (MeV) & & & & & \\
\hline
\hline
 &          & & & & & \\
1&$\bar 3$&1&0$(1^+)$&2455.91&0&$D_{s1}(2460)$ \\
\cline{3-7}
 &2432.63 &0&$\oh(1^+)$&2311.24&-115.68&$D_1(2430)$ \\
\cline{2-7}
 & 6   &1&1$(1^+)$&2529.30&-238.56& (?) \\
\cline{3-7}
& 2532.57&0&$\oh(1^+)$&Cusp (2607)&Broad& (?) \\
\cline{3-7}
& -i199.36&-1&0$(1^+)$&Cusp (2503)&Broad&(?) \\
\cline{2-7}
&  &1&0$(1^+)$&2573.62&-0.07&$D_{s1}(2536)$ \\
&  $\bar 3$   &  &      &     & [-0.07] & \\
\cline{3-7}
&2535.07&0&$\oh(1^+)$&2526.47&-0.08&$D_1(2420)$ \\
& -i0.08& &     &      & [-13]     & \\
\cline{2-7}
& 6 & 1&1$(1^+)$& 2756.52&-32.95&(?) \\
&  &   &            &         & [cusp] & \\
\cline{3-7}
&Cusp (2700)&0&$\oh(1^+)$&2750.22&-99.91&(?) \\
&           &     &    &  &[-101]& \\
\cline{3-7}
&Narrow&-1&0$(1^+)$&2756.08&-2.15&(?) \\
&      &     &       &   &[-92] & \\
\hline
0&1 &0&0$^-(1^{+-})$&925.12&-24.61&$h_1(1170)$ \\
 &1055.77& & & & & \\
\cline{2-7}
& 8&1&$\oh(1^+)$&1101.72&-56.27&$K_1(1270)$ \\
\cline{3-7}
&1161.06&0&1$^+(1^{+-})$&1230.15&-47.02&$b_1(1235)$ \\
\cline{4-7}
& & &0$^-(1^{+-})$&1213.00&-5.67&$h_1(1380)$ \\
\cline{2-7}
&1&0&0$^+(1^{++})$&3837.57&-0.00&$X(3872)$\\
&3867.59& & & & & \\
\cline{2-7}
&8&1&$\oh(1^+)$&1213.20&-0.89&$K_1(1270)$ \\
\cline{3-7}
&1161.37&0&1$^-(1^{++})$&1012.95&-89.77&$a_1(1260)$ \\
\cline{4-7}
& &  &0$^+(1^{++})$&1292.96&0&$f_1(1285)$ \\
\cline{2-7}
&1&0&0$^-(1^{+-})$&3840.69&-1.60&(?)\\
&3864.62& & & & & \\
&-i0.00& & & & & \\
\hline
\end{tabular}
\caption{Pole positions for the model. The column Irrep shows the results in the $SU(3)$ limit. The results in brackets for the $Im\sqrt{s}$ are obtained taking into account the finite width of the $\rho$ and $K^*$ mesons.} \label{resultsx1}
\end{center}
\end{table}

We will now discuss separately the particle identification in each sector.

\subsection{C=1,S=1,I=1}

In contrast with the scalar resonances where the sextet state became very broad [\ref{gamer}], the axial sextets are narrower, hence easier to detect experimentally. One should note also that these states are truly exotics since quark models cannot generate $q\bar q$ pairs with such quantum numbers. We found two poles in this sector at positions (2529.30-i238.56) MeV and (2756.52-i32.95) MeV.

The couplings, $g_i$, of the poles to each channel $i$ have been calculated from the residues of each pole. Close to the pole position one can write:

\begin{eqnarray}
T_{ij}&\cong&{g_i g_j \over s-s_{pole}}
\end{eqnarray}

Table \ref{sexi1} shows the results of $g_i$ for the poles in this sector. With the couplings it is possible to do a rough estimate of the partial decay widths for the resonances and thus identify the channels with largest contribution to the width in order to motivate experimental searches in this direction. At tree level one has:

\begin{eqnarray}
\Gamma_{A\rightarrow PV}&=&{|g_i|^2\over8\pi M_A^2}p
\end{eqnarray}
where p is the center of mass three-momentum of the two particles in the final state.

\begin{table}
\begin{center}
\begin{tabular}{c||c|c}
\hline
Channel & (2529.30-i238.56) MeV & (2756.52-i32.95) MeV \\
& $|g_i|$ (GeV) & $|g_i|$ (GeV) \\
\hline
\hline
$\pi D_s^*$&9.13&2.54 \\
\hline
$D_s\rho$&2.18&9.26\\
\hline
$DK^*$ & 1.94 & 11.02 \\
\hline
$KD^*$&8.20&2.61 \\
\hline
\end{tabular}
\caption{Residues for the C=1,S=1,I=1 sector}\label{sexi1}
\end{center}
\end{table}

 For the pole at (2756.52-i32.95) MeV the estimate reads:

\begin{center}
${\Gamma_{D_s \rho}\over\Gamma_{\pi D_s^*}}\sim 3.3$ and ${\Gamma_{D_s \rho}\over\Gamma_{K D^*}}\sim 3.6$
\end{center}

 While for the pole at (2529.30-i238.56) MeV on has:

\begin{center}
${\Gamma_{\pi D_s^*}\over\Gamma_{K D^*}}\sim 3.1$
\end{center}

 The large coupling of the lighter state to $\pi D_s^*$ and $KD^*$, or the heavier one to $D_s\rho$ and $DK^*$ make these states qualify as roughly quasi-bound states of these channels respectively. Note that they separate two basic configurations: heavy vector-light pseudo-scalar and heavy pseudo-scalar-light vector.
 
When taking into account the finite $\rho$ and $K^*$ widths the resonance at (2756.52-i32.95) becomes a little bit higher in mass, crossing the $KD^*$ thresholds and disappearing as a pole.

\subsection{C=1,S=1,I=0}

The two poles found in this sector have the proper quantum numbers to be identified with the two $D_{s1}$ resonances. The first pole appears as an exact bound state at 2455 MeV and we identified it with the $D_{s1}(2460)$ state. Experimentally the main hadronic decay channel for this resonance is $D_s^*\pi$ which is an isospin violating decay and therefore not taken into account by our model. Other decays for this resonance are three body decays or electromagnetic ones, which are also not included in our framework.

The other pole appears at (2573.62-i0.07) MeV and couples mainly to the $DK^*$ and $D_s\omega$ channels. The only open channel for it to decay is the $KD^*$ channel but, because of the dynamics of the interaction, this resonance barely couples to it. This explains the small width of this resonance, 140 KeV, despite the 70 MeV phase space available for it to decay. We identify this pole with the $D_{s1}(2536)$ which is also observed in the decay channel $KD^*$ with a small width ($\Gamma<2.3$ MeV [\ref{pdg}]).

Table \ref{ds1} shows the absolute value of the couplings $|g_i|$ for each channel for the two poles in this sector.

 Once more we see that the lighter state couples strongly to $KD^*$ and $\eta D_s^*$ while the second one couples strongly to $DK^*$ and $D_s\omega$. Hence the decoupling into two families of heavy vector-light pseudo-scalar and light vector-heavy pseudo-scalar shows up in this sector too.

\begin{table}
\begin{center}
\begin{tabular}{c||c|c}
\hline
Channel & 2455.91 MeV & (2573.62-i0.07) MeV \\
& $|g_i|$ (GeV) & $|g_i|$ (GeV) \\
\hline
\hline
$DK^*$ & 0.54 & 13.96 \\
\hline
$KD^*$&9.74&0.30 \\
\hline
$\eta D_s^*$&6.00&0.18 \\
\hline
$D_s\omega$&0.51&7.95 \\
\hline
$\eta_c D_s^*$&0.02&0.05 \\
\hline
$D_s J/\psi$&0.54&0.00 \\
\hline
\end{tabular}
\caption{Residues for the C=1,S=1,I=0 sector}\label{ds1}
\end{center}
\end{table}

This sector has one resonance with a strong coupling to the $D_s \omega$ channel ($\omega=\omega_8$ here). Hence, this is one case where the $\phi-\omega$ mixing can be relevant. Since we saw that the singlet ($\omega_1$) does not give any contribution to the Lagrangian, the explicit introduction of the $\omega$ and $\phi$ states can be done by substituting:

\begin{equation}
\omega_8=\squt \omega-\sqdt\phi
\end{equation}

This splitting into two fields introduces a new column and a new row in the second table of appendix A.1., by simply multiplying the fourth column and row by $\squt$ and introducing a $D_s\phi$ column and row with weights $-\sqdt$ those of the original $D_s\omega_8$. When we look now for poles we obtain the results in Table \ref{withphi}.

Comparing Tables \ref{ds1} and \ref{withphi}, we can see that in the case where the coupling of the resonance to the $D_s\omega_8$ channel is weak (first resonance) the effects of the mixing are very small in the energy and couplings of the resonance. In the case of the second resonance, where the coupling to the $D_s\omega_8$ channel was large, the effects of the mixing are more visible. There is a shift of the mass of about 25 MeV, which is well within our theoretical uncertainties. This effect can be considered an upper bound for all other cases, since we have chosen the resonance with strongest coupling to $\omega_8$.

It is also interesting to see that the sums of the squares of the couplings to $\omega$ and $\phi$ are close to the square of that to $\omega_8$, indicating a redistribution of the strength of the coupling to $\omega_8$ between $\omega$ and $\phi$.

\begin{table}
\begin{center}
\begin{tabular}{c||c|c}
\hline
Channel & 2455.97 MeV & (2597.44-i0.06) MeV \\
& $|g_i|$ (GeV) & $|g_i|$ (GeV) \\
\hline
\hline
$DK^*$ & 0.41 & 13.86 \\
\hline
$KD^*$&9.82&0.27 \\
\hline
$\eta D_s^*$&6.06&0.09 \\
\hline
$D_s\omega$&0.27&5.12 \\
\hline
$D_s\phi$&0.36&7.06 \\
\hline
$\eta_c D_s^*$&0.02&0.01 \\
\hline
$D_s J/\psi$&0.55&0.00 \\
\hline
\end{tabular}
\caption{Residues for the C=1,S=1,I=0 sector when introducing the $D_s\phi$ channel}\label{withphi}
\end{center}
\end{table}

The widths of the light vector mesons have no significant effects over the resonances generated in this sector, because the mass of the resonances are far away from the threshold of the $DK^*$ channel.

\subsection{C=1,S=0,I=$\oh$}

Here the companions of the two anti-triplets and the two sextets should be found. Note that when we refer to the $SU(3)$ multiplet we are talking about the case when one has $SU(3)$ symmetry. This correspond to $x=0$ in the pole trajectories. At $x=1$, since $SU(3)$ symmetry is broken, the physical states mix the $SU(3)$ multiplets. Yet, the study of the trajectories allows us to trace back any pole to its origin in the $SU(3)$ symmetric case, and we have used this information for the classification of states in Table \ref{resultsx1}.

The anti-triplet companion of the pole for the $D_{s1}(2460)$ is the pole located at (2311.24-i115.68) MeV that we identify with the $D_1(2430)$ because of its naturally large width, since it is strongly coupled to the $\pi D^*$ channel into which it is free to decay. On the other hand the pole at (2526.47-i0.08) MeV, companion of the one identified with the $D_{s1}(2536)$, has its coupling to the $\pi D^*$ channel strongly suppressed and therefore has a very narrow width. Because of this unnatural narrow width we are tempted to identify it with the $D_1(2420)$ although the mass of our dynamically generated resonance is around 100 MeV off the experimental value for this state. Moreover, when considering the finite widths of the $\rho$ and $K^*$ mesons, this pole gets a larger width, its imaginary part goes to $-13$ MeV, implying a width of about 26 MeV, in fair agreement with experiment.

 As for the sextets, one of the poles becomes a broad cusp at the $\bar K D_s^*$ threshold as one gradually breaks $SU(3)$ symmetry through the parameter $x$, and the other pole emerges from a cusp into a pole at (2750.22-i99.91) MeV. The channel to which it is most strongly coupled is closed, the $D_s \bar K^*$, but it also has sensitive couplings to all channels into which it is allowed to decay. The consideration of the finite width of the vector mesons has very small effect over this resonance ($\sim$2 MeV increase in its width).

The couplings of the poles in this sector to the channels are given in Table \ref{d0ss}.

\begin{table}
\begin{center}
\begin{tabular}{c||c|c|c}
\hline
Channel & (2311.24-i115.68) MeV &(2526.47-i0.08) MeV & (2750.22-i99.91) MeV \\
& $|g_i|$ (GeV) & $|g_i|$ (GeV)& $|g_i|$ (GeV) \\
\hline
\hline
$\pi D^*$&9.84&0.24&2.15 \\
\hline
$D\rho$&0.89&12.13&3.82 \\
\hline
$\bar K D_s^*$&5.21&0.59&2.38 \\
\hline
$D_s \bar K^*$&0.09&7.89&13.11 \\
\hline
$\eta D^*$&0.68&0.56&2.43 \\
\hline
$D\omega$&0.61&1.58&7.47 \\
\hline
$\eta_c D^*$&0.06&0.02&0.22 \\
\hline
$D J/\psi$&1.27&0.02&0.01 \\
\hline
\end{tabular}
\caption{Residues for the C=1,S=0,I=$\oh$ sector}\label{d0ss}
\end{center}
\end{table}

 As in the former cases, the states are clearly separated into the heavy vector-light pseudo-scalar and light vector-heavy pseudo-scalar sectors.

\subsection{C=1,S=-1,I=0}

The two remaining exotic members of the sextet should be found in this sector. One of them becomes a broad cusp at the $\bar K D^*$ threshold when $x=1$ while the other one is a narrow resonance with pole position (2756.08-i2.15) MeV. The couplings of this pole are given in Table \ref{sexs-1}.

\begin{table}
\begin{center}
\begin{tabular}{c||c}
\hline
Channel & (2756.08-i2.15) MeV  \\
& $|g_i|$ (GeV)  \\
\hline
\hline
$D\bar K^*$ & 5.66  \\
\hline
$\bar KD^*$&1.42\\
\hline
\end{tabular}
\caption{Residues for the C=1,S=-1,I=0 sector}\label{sexs-1}
\end{center}
\end{table}

When taking into account the 50 MeV width of the $K^*$ meson, this resonance gets a much bigger width, of the order of 180 MeV. In this case, and in all other sectors where the effect of the finite width of the vector mesons were taken into account, the only significant effect one could observe was over the width of the resonance. The mass of the resonances were affected in less than 0.5 \% and the important couplings in less than 5\%.

\subsection{C=0,S=1,I=$\oh$}

 Two poles are found here coming from the two octets in the scattering of the low lying pseudo-scalar with the light vector mesons. In principle one could be tempted to assign these two poles to the two axial kaons from PDG [\ref{pdg}], but the mass of one of these, the $K_1(1400)$ is about 200-300 MeV off the pole positions we found and its width is much smaller than that. With this in mind we followed the interpretation of Roca [\ref{roca}] that the $K_1(1270)$ should have a two pole structure.

 The couplings of the two poles to the different channels are in Table \ref{kaonscoup}.

\begin{table}
\begin{center}
\begin{tabular}{c||c|c}
\hline
Channel & (1101.72-i56.27) MeV & (1213.20-i0.89) MeV \\
& $|g_i|$ (GeV) & $|g_i|$ (GeV) \\
\hline
\hline
$\pi K^*$&4.48&0.51 \\
\hline
$K\rho$&1.57&5.15 \\
\hline
$\eta K^*$&0.36&3.55 \\
\hline
$K\omega$&3.02&1.42 \\
\hline
$\bar D D_s^*$&0.78&0.16 \\
\hline
$D_s \bar D^*$&0.05&0.48 \\
\hline
$K J/\psi$&0.08&0.02 \\
\hline
$\eta_c K^*$&0.03&0.02 \\
\hline
\end{tabular}
\caption{Residues for the C=0,S=1,I=$\oh$ sector}\label{kaonscoup}
\end{center}
\end{table}

 This sector is explained in more detail in [\ref{roca}]. The novelty here is that, in spite of including now the heavy channels, the results are basically unaltered compared to those of [\ref{roca}] where only the light sector was used. This indicates a very weak mixing of the heavy and light sectors.

 Concerning the two $K_1$ states it is also opportune to mention that in [\ref{geng}] some experimental information was reanalyzed giving strong support to the existence of these two states.

\subsection{C=0,S=0,I=1}

In this sector too there are two poles coming from the two octets but, since this is the non-strange sector, this two states have defined G-parity and therefore cannot mix. 

The pole with positive G-parity we associate with the $b_1(1235)$ resonance. The small discrepancy between the experimental width and the value found from our theoretical model is explained since, experimentally, some decay channels of this resonance are three or four body decays while our model contemplates just two body hadronic decays.

The negative G-parity pole should be identified with the $a_1(1260)$ but here the model gives a worse description of the resonance, the mass of the pole is smaller than expected although the huge width of the resonance makes this a minor problem. Also the width found within the model is very large, of the order of magnitude of the experimental one which is estimated with large errors. Again one should note that an important fraction of the width of this resonance could be due to many body decays not included in the present model.

The couplings of the resonances to the channels are given in Table \ref{a1b1} and they are very similar to those found in [\ref{roca}]. There the $\omega_8$ was substituted in terms of $\omega$ and $\phi$ and the sum of $|g|^2$ for $\pi\omega$ and $\pi\phi$ is similar to the $|g|^2$ for $\pi\omega_8$ of our calculation.

\begin{table}
\begin{center}
\begin{tabular}{c||c|c}
\hline
Channel & (1230.15-i47.02) MeV & (1012.95-i89.77) MeV \\
& $|g_i|$ (GeV) & $|g_i|$ (GeV) \\
\hline
\hline
$\pi \rho$&-&4.49 \\
\hline
$K\bar K^*\pm c.c.$&6.56&2.42 \\
\hline
$\pi\omega$&3.07&-\\
\hline
$\eta\rho$&2.90&-\\
\hline
$D\bar D^*\pm c.c.$&0.45&0.88\\
\hline
$\eta_c\rho$&0.03&-\\
\hline
$\pi J/\psi$&0.04&-\\
\hline
\end{tabular}
\caption{Residues for the C=0,S=0,I=1 sector}\label{a1b1}
\end{center}
\end{table}

\subsection{C=0,S=0,I=0}

Five poles are found in this sector. Tree have negative charge conjugation parity and two of them positive C-parity. In the light sector the positive C-parity pole is associated with the $f_1(1285)$, it appears in our model as a truly bound state, as it should, since none of its observed decay channels is a pseudo-scalar vector meson one, the possible decay channels within the model. The results obtained here and in the other two sectors for the light axial resonances without $\omega-\phi$ mixing are very similar to those obtained in [\ref{roca}] where the mixing was explicitly taken into account thus, corroborating the moderate effects of the mixing found for the charmed sector.

The heavy singlet with positive C-parity obtained at 3837 MeV is a good candidate to be associated with the controversial state $X(3872)$. In this case this state is interpreted as being mainly a mixed molecule of $D\bar D^*+c.c.$ and $D_s\bar D_s^*-c.c.$, its only possible decay channel within the model being the $K\bar K^*+c.c.$ which is highly suppressed. In Table \ref{poscp} the couplings of the two poles are presented. We can see there the strong decoupling of the heavy and light sectors.

The low lying negative C-parity resonances can be associated with the two $h_1$ resonances. The singlet at (925.12-i24.61) MeV we identify with the $h_1(1170)$ and, since we get it with a lower mass, our width is much smaller than the experimental one because our state has less phase space for decay. With the octet pole at (1213.00-i5.67) MeV the same thing happens, and we associate it with the $h_1(1380)$ despite the smaller mass and width compared with experimental values.

In the heavy sector we find another state at 3840 MeV and negative C-parity. The decays of the $X(3872)$, reported in [\ref{abe2}], into $\gamma J/\psi$ and $\omega J/\psi$ indicate that the C-parity of this state is positive. The decay into $\pi^+\pi^-J/\psi$ reported in [\ref{expx1}] assuming that the $\pi^+\pi^-$ comes from a $\rho$ state would give the same C-parity but would imply isospin breaking. The large branching fraction

\begin{eqnarray}
{B(X\rightarrow\pi^+\pi^-\pi^0 J/\psi) \over B(X\rightarrow\pi^+\pi^- J/\psi)}&=&1.0\pm0.4\pm0.3 \label{branch}
\end{eqnarray}
reported in [\ref{abe2}] indicates a massive violation of G-parity and hence isospin if one has only one $X$ particle.

There is a more appealing explanation for eq. (\ref{branch}) if one had two $X(3872)$ states with different G-parity and correspondingly C-parity. Should the $\pi^+\pi^-$ in the denominator of eq. (\ref{branch}) correspond to an I=0 state one would not have to invoke isospin violation, but instead the existence of a negative G-parity (and hence C-parity) state. This would imply that there is strength of these events in the $\sigma$ region of the $\pi\pi$ invariant mass, and this seems to be the case as reported in [\ref{expx4}], although the statistics is low. This latter scenario would fit with our predictions of two states nearly degenerate with opposite C-parity. A similar argumentation has been made in [\ref{terasaki}] to justify the existence of two degenerate $X(3872)$ states with different C-parity.

The difference of 35 MeV in the binding energy between our model and experiment ($\sim$ 1\% difference) is perfectly acceptable for a theoretical model that looks at the whole spectrum of axial vector mesons in a broad range of masses with only two free parameters ($\alpha_H$, $\alpha_L$). Yet, we can do some fine tuning to get a mass like in experiment by changing the subtraction constants in the loops. In this case we can take $\alpha_H=-1.30$ (from $-1.55$ before) and then we find the mass of the positive C-parity pole at 3872.67 MeV and simultaneously the state with negative C-parity just disappears as a pole, by crossing the $\bar D D^*$ threshold, and leads to a marked cusp structure in the $\bar DD^*-c.c.$ amplitude. Indeed, changing $\alpha_H$ to slightly more negative values we regain the pole just below this threshold.

Recently a new possible state at 3875 MeV was reported at Belle [\ref{17han}] decaying into $\bar{D_0}D_0^*$. The 3 MeV difference of this state with the $X(3872)$ is precisely the difference of masses between the positive and negative C-parity states that we obtain. It is thus tempting to associate to the new $X(3875)$ state our negative C-parity state. Such scenario is not excluded by the data as we show in the analysis below.

 In the decay of a B particle to $K\pi^+\pi^-J/\psi$ we have, defining $E=M_{inv}(\pi^+\pi^- J/\psi)-M_{\bar D^0 D^{0*}}$:

\begin{eqnarray}
\frac{dBr_+}{dE}&\propto&|T(\bar D^0D^{0*}+c.c.\rightarrow\bar D^0D^{0*}+c.c.)|^2\Gamma(X\rightarrow\pi^+\pi^-J/\psi)
\end{eqnarray}
which is approximately proportional to $|T|^2$ in a few MeV range of E around $M_{\bar D^0 D^{0*}}$ since $\Gamma(X \rightarrow\pi^+\pi^-J/\psi)$ barely changes in this range of energies given the relatively large phase space for this decay. On the other hand for the B particle decaying into $K\bar D^0 D^{0*}$ we have (now $E=M_{inv}(\bar D^0 D^{0*})-M_{\bar D^0 D^{0*}}$):

\begin{eqnarray}
\frac{dBr_-}{dE}&\propto&|T(\bar D^0D^{0*}-c.c.\rightarrow\bar D^0D^{0*}-c.c.)|^2\Gamma(X'\rightarrow\bar D^0D^{0*})
\end{eqnarray}
which is approximately proportional to $|T|^2 p$, where p is the three-momentum of the $\bar D$ particle in the center of mass frame of the $\bar DD^*$ system, since now $\Gamma(X'\rightarrow \bar D^0D^{0*})$ is proportional to this three-momentum.

\begin{figure}
\begin{tabular}{cc}
\includegraphics[width=7cm]{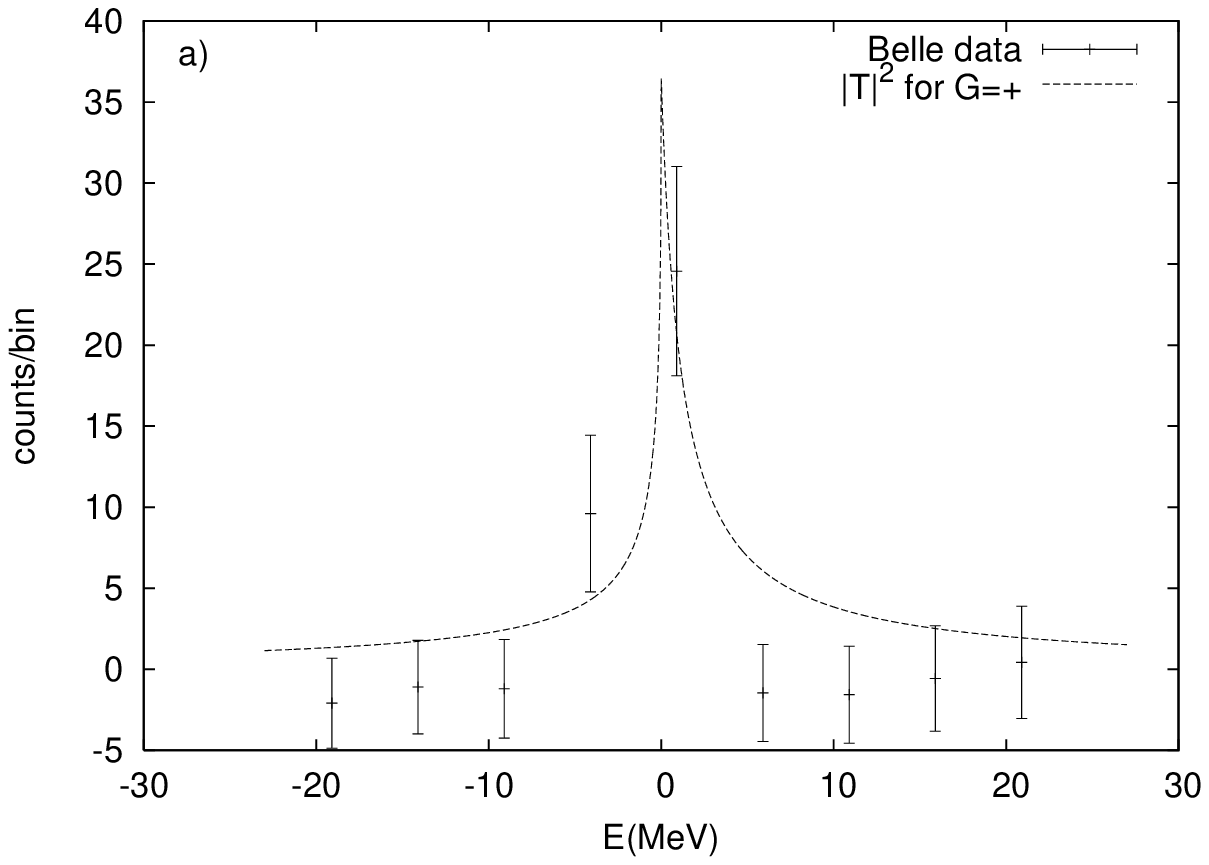} & \includegraphics[width=7cm]{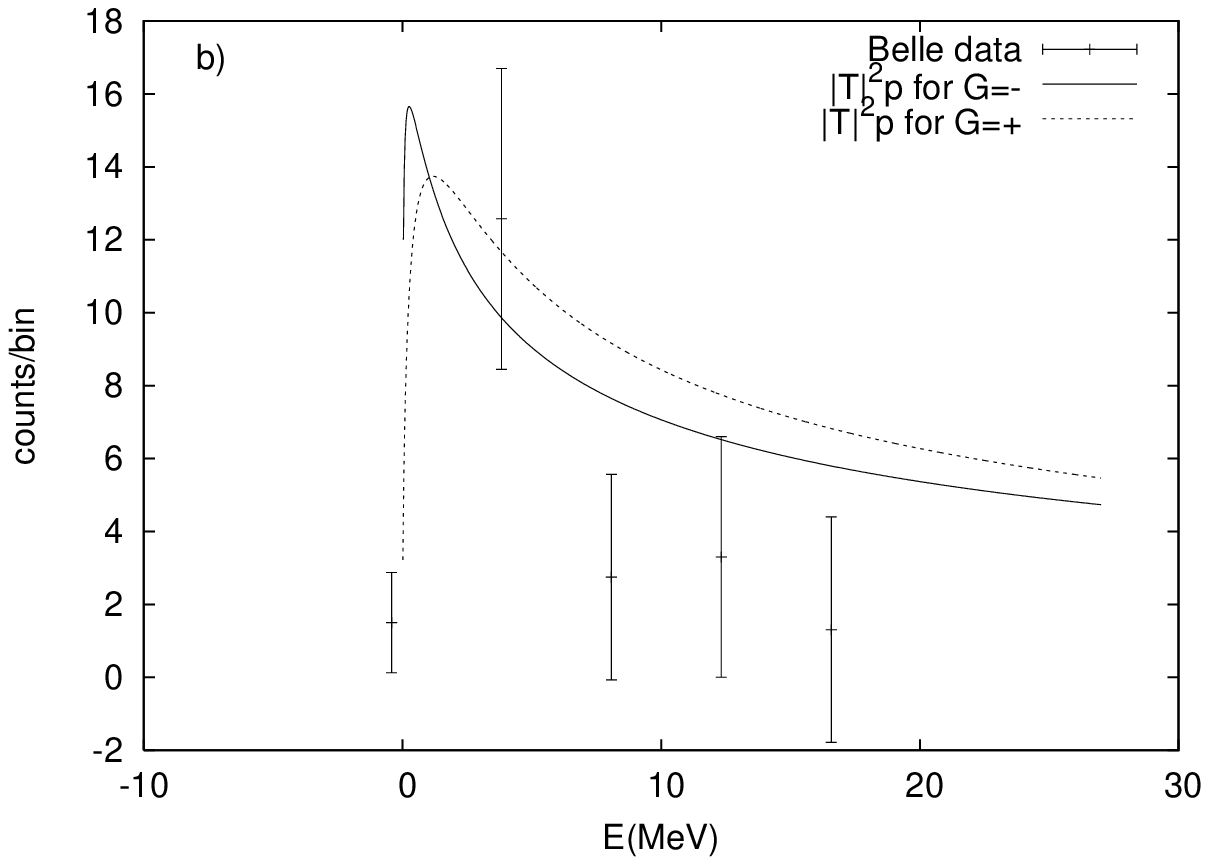} \\
\end{tabular}
\caption{a)$|T|^2$ for the positive G-parity state in the $D\bar D^*$ channel compared with the Belle Data (in this plot $\alpha_H=-1.23$) b) $|T|^2 p$ for both G-parity states in the $D\bar D^*$ channel compared with the Belle Data (in this plot $\alpha_H=-1.30$ for the G=- state)} \label{belle}
\end{figure}

 In fig \ref{belle} we plot $\frac{dBr_+}{dE}$ and $\frac{dBr_-}{dE}$ as a function of $E$ and we compare our results with the experimental data from Belle [\ref{17han}] [\ref{23han}]. We can see that in the case of the positive C-parity state the $\pi^+\pi^-J/\psi$ distribution is very sharp, while the distribution of the invariant mass of $\bar D^0 D^{0*}$ leads to a strong enhancement of the distribution around the $\bar D^0 D^{0*}$ threshold, in both cases in fair agreement with experiment.

 We should note that an empirical analysis of the data in a recent paper [\ref{hanh}], taking only one resonance, produces a similar behavior assuming that the resonance couples very strongly to $D^0\bar  D^{0*}$, as it is also our case (see couplings in Tables \ref{poscp}-\ref{negcp}). In view of the results of [\ref{hanh}] it is also instructive to see what our model would give assuming that the resonance with positive C-parity is responsible for the two distributions in fig. \ref{belle}. We show the result obtained for $|T|^2p$ with the positive C-parity resonance in figure \ref{belle} (b) with dotted line. As we can see, the shape and strength obtained in this case is not very different from that of the negative C-parity resonance. Hence, this alternative scenario, which would correspond to the one in [\ref{hanh}], is not ruled out by these combined data.
 
As mention above the ratio of eq. (\ref{branch}) and the possible strength seen for the $B(X\rightarrow\pi^+\pi^- J/\psi)$ in the region of invariant $\pi\pi$ masses around $m_\sigma$, is so far the strongest indications favoring the two C-parity states.

\begin{table}
\begin{center}
\begin{tabular}{c||c|c}
\hline
Channel & 1292.96 MeV & (3837.57-i0.00) MeV \\
& $|g_i|$ (GeV) & $|g_i|$ (GeV) \\
\hline
\hline
$\squd(D\bar D^*+ c.c.)$&0.15&13.61\\
\hline
$\squd(D_s \bar D_s^*- c.c.)$&0.54&10.58 \\
\hline
$\squd(K\bar K^*+ c.c.)$&7.15&0.03 \\
\hline
\end{tabular}
\caption{Residues for the C=0,S=0,I$^P$=0$^+$ sector}\label{poscp}
\end{center}
\end{table}

\begin{table}
\begin{center}
\begin{tabular}{c||c|c|c}
\hline
Channel & (925.12-i24.61) MeV & (1213.00-i5.67) MeV& (3840.69-i1.60) MeV \\
& $|g_i|$ (GeV) & $|g_i|$ (GeV)& $|g_i|$ (GeV) \\
\hline
\hline
$\pi\rho$&3.65&1.07&0.03 \\
\hline
$\eta\omega$&0.10&4.19&0.00 \\
\hline
$\squd(D\bar D^*- c.c.)$&3.93&1.03&13.44 \\
\hline
$\eta_c\omega$&0.00&0.04&1.25 \\
\hline
$\eta J/\psi$&0.01&0.06&1.20 \\
\hline
$\squd(D_s \bar D_s^*+ c.c.)$&2.31&1.26&9.96 \\
\hline
$\squd(K\bar K^*- c.c.)$&0.73&6.14&0.01 \\
\hline
$\eta_c J/\psi$&1.08&0.34&1.99 \\
\hline
\end{tabular}
\caption{Residues for the C=0,S=0,I$^P$=0$^-$ sector}\label{negcp}
\end{center}
\end{table}

\section{Overview and Conclusion}

 We studied the dynamical generation of axial resonances by looking for the poles in the scattering T-matrix of pseudo-scalars with vector mesons. For the interaction Lagrangian we first constructed a $SU(4)$ flavor symmetrical Lagrangian for the interaction of the 15-plet of pseudo-scalar mesons with the 15-plet of vector mesons. The symmetry was broken down to $SU(3)$ by suppressing exchanges of heavy vector mesons in the implicit Weinberg-Tomozawa term, following a prescription developed in a previous paper [\ref{gamer}]. From the Lagrangian, tree level amplitudes were evaluated, projected in s-wave and collected in a matrix for the various channels. This matrix was used as the potential, transformed to an isospin basis, in the Bethe-Salpeter equation, which provides the unitarized amplitudes between the channels.

 The poles generated within the model can be associated with the various axial resonances listed by the Particle Data Group [\ref{pdg}], and also many new resonances are predicted: three broad ones in the mass range between 2.5 and 2.6 GeV, and three narrower ones around 2.75 GeV, these resonances belong to two $SU(3)$ sextets, they have, therefore, exotic quantum numbers and they have not yet been experimentally observed. We summarize these states with their quantum numbers in Table \ref{newres}. 

\begin{table}[h]
\begin{center}
\begin{tabular}{c|c|c|c|c|c}
\hline
C& S&I$^G(J^{PC})$& Re($\sqrt{s}$) (MeV)& Im($\sqrt{s}$) (MeV)&Channel\\
\hline
\hline
1  &1&1$(1^+)$&2529.30&-238.56& $\pi D_s^*$, $KD^*$ \\
\hline
1& 0&$\oh(1^+)$&Cusp (2607)&Broad& - \\
\hline
1& -1&0$(1^+)$&Cusp (2503)&Broad& - \\
\hline
1&  1&1$(1^+)$& 2756.52&-32.95 [cusp]&$D_s\rho$, $DK^*$ \\
\hline
1&0&$\oh(1^+)$&2750.22&-99.91 [-101]&$D\omega$ \\
\hline
1&-1&0$(1^+)$&2756.08&-2.15 [-92]& $D\bar K^*$ \\
\hline
0&0&0$^-(1^{+-})$&3840.69&-1.60&$\eta_c\omega$, $\eta J/\psi$\\
\hline
\end{tabular}
\caption{List of predicted and not yet observed resonances with quantum numbers and the open channels to which they couple most strongly. The results in brackets for the $Im\sqrt{s}$ are obtained taking into account the finite width of the $\rho$ and $K^*$ mesons.} \label{newres}
\end{center}
\end{table}

Remaining discrepancies between our model and experiment can be attributed to possible many body decays of these objects or possible higher order terms that could be included in the Lagrangian. The size of the discrepancies, typical of any successful hadronic model describing hadronic spectra, can be used to estimate the uncertainties in the predictions for the new states that we obtain.

 Some states obtained here have been also studied before in a similar framework, but with more restricted coupled channels space. The similarity of the results reinforces these findings. Many other states are reported here for the first time within this unitary coupled channel framework. In the light scalar sector the poles and couplings found within our approach coincide very well with the ones found by Roca [\ref{roca}]. This happens because, despite the enlarged coupled channel space including heavy mesons, this new sector couples very weakly with the light one. In the open-charm sector the poles found for the lightest anti-triplet coincide with the results found by Kolomeitsev [\ref{lutz1}] and Guo [\ref{chiang1}]. As already happened for the scalar resonances, the poles found within our model for the sextet state in this sector have broader widths because of the use, in our model, of a different meson decay constant for heavy mesons. Besides, our model allows also for the inclusion of channels with heavy pseudo-scalar mesons interacting with light vector ones, as a result of which our model generates a richer spectrum, with poles for an extra anti-triplet and an extra sextet. In the charmed sector, C=1, we find six resonances not yet observed. Three of them are either too broad, or they degenerate into cusps as $SU(3)$ is gradually broken. However, three of them remain sufficiently narrow, such that they could in principle be detected.

 Moreover our Lagrangian incorporates the hidden-charm sector and an attractive interaction in the $3\otimes\bar3^*$ and $\bar 3\otimes3^*$ charmed mesons is responsible for the generation of two resonances. One of them can be associated with the new $X(3872)$ state. The other one appears as a strong cusp in our case, or a resonance with slightly increased attraction, and could be associated to the peak structure found recently around the $\bar D^0D^{0*}$ threshold at Belle, although this structure can also be explained in terms of only one state. On the other hand, the ratio of partial decay width into two or three pions and $J/\psi$ (eq. (\ref{branch})) provides strong support for the existence of two states with different C-parity.

 The agreement of the theory with data for the known resonances, together with the success of the theory for the charmed scalar mesons [\ref{gamer}], gives us confidence on these new predicted states, such as to strongly suggest their experimental search.

\section{Acknowledgements}  
We would like to thank prof. K. Terasaki for some useful discussions..
This work is partly supported by DGICYT contract
number BFM2003-00856 and the Generalitat Valenciana. This research is  part of the EU 
Integrated
Infrastructure Initiative Hadron Physics Project under contract number
RII3-CT-2004-506078.

\newpage

\newpage

\appendix

\section{The $\xi$ Coefficients}

\subsection{Open-Charm Sector (C=1)}

\center{\underline{S=1, I$^G(J^{PC})$=$1(1^+)$}}

\begin{tabular}{c|cccc}
Channels & $\pi D^*_s$ & $D_s\rho$ & $K D^*$ & $D K^*$ \\
\hline
 $\pi D^*_s$ &0 & 0 & -1 & -$\gamma$  \\
 $D_s\rho$ &0 & 0 & -$\gamma$  & -1 \\
 $K D^*$ &-1 & -$\gamma$ & 0 & 0 \\
 $D K^*$&$-\gamma$  & -1 & 0 & 0
\end{tabular}

\hbox{    }

\hbox{    }

\center{\underline{S=1, I$^G(J^{PC})$=$0(1^+)$}}

\begin{tabular}{c|cccccc}
Channels & $DK^*$ & $KD^*$ & $\eta D_s^*$ & $D_s \omega$ & $\eta_c D_s^*$ & $D J/\psi$ \\
\hline
 $DK^*$ &-2 & 0 & -$\frac{\gamma }{\sqrt{3}}$ & -$\sqrt{3}$ & -2
  $ \sqrt{\frac{2}{3}} \gamma$  & 0 \\
 $KD^*$ &0 & -2 & $\sqrt{3}$ &$ \frac{\gamma }{\sqrt{3}} $& 0 & 2
  $ \sqrt{\frac{2}{3}} \gamma $ \\
$\eta D_s^*$ &$- \frac{\gamma }{\sqrt{3}} $&$ \sqrt{3} $& 0 & $\frac{2
   \gamma }{3}$ & 0 &$- \frac{2 \sqrt{2} \gamma }{3} $\\
 $D_s \omega$ &-$\sqrt{3}$ & $\frac{\gamma }{\sqrt{3}}$ &$ \frac{2 \gamma
   }{3} $& 0 & -$\frac{2 \sqrt{2} \gamma }{3}$ & 0 \\
 $\eta_c D_s^*$ &$-2 \sqrt{\frac{2}{3}} \gamma $ & 0 & 0 & $-\frac{2 \sqrt{2}
   \gamma }{3}$ & 0 & $\frac{4 \gamma }{3}$ \\
 $D J/\psi$&0 & $2 \sqrt{\frac{2}{3}} \gamma $ &-$ \frac{2 \sqrt{2}
   \gamma }{3}$ & 0 &$ \frac{4 \gamma }{3}$ & 0
\end{tabular}

\hbox{    }

\hbox{    }

\center{\underline{S=0, I$^G(J^{PC})$=$\oh(1^+)$}}

\begin{tabular}{c|cccccccc}
 Channels & $\pi D^*$ & $D\rho$&$\bar{K}D_s^*$&$D_s\bar{K^*}$&
$\eta D$&$D\omega$&$\eta_c D^*$&$D J/\psi$ \\
\hline
 $\pi D^*$ &-2 & $\frac{\gamma }{2} $&$ \sqrt{\frac{3}{2}}$ & 0 & 0 &
  -$ \frac{\gamma }{2}$ & 0 & -$\sqrt{2} \gamma $ \\
  $D\rho$&$\frac{\gamma }{2}$ & -2 & 0 & -$\sqrt{\frac{3}{2}}$ &
   $\frac{\gamma }{2}$ & 0 & $\sqrt{2} \gamma$  & 0 \\
 $\bar{K}D_s^*$&$\sqrt{\frac{3}{2}}$ & 0 & -1 & 0 & -$\sqrt{\frac{3}{2}}$ &
  -$ \sqrt{\frac{2}{3}} \gamma $ & 0 &$ \frac{2 \gamma
   }{\sqrt{3}}$ \\
 $D_s\bar{K^*}$&0 &$- \sqrt{\frac{3}{2}} $& 0 & -1 &- $\sqrt{\frac{2}{3}}
   \gamma $ & -$\sqrt{\frac{3}{2}} $& $\frac{2 \gamma
   }{\sqrt{3}}$ & 0 \\
$\eta D$& 0 & $\frac{\gamma }{2}$ & -$\sqrt{\frac{3}{2}}$ &
  - $\sqrt{\frac{2}{3}} \gamma$  & 0 &$ \frac{\gamma }{6}$ &
   0 & $\frac{\sqrt{2} \gamma }{3}$ \\
 $D\omega$&-$\frac{\gamma }{2}$ & 0 &-$\sqrt{\frac{2}{3}} \gamma $ &
   $-\sqrt{\frac{3}{2}}$&$ \frac{\gamma }{6}$ & 0 &
   $\frac{\sqrt{2} \gamma }{3}$ & 0 \\
 $\eta_c D^*$&0 &$ \sqrt{2} \gamma $ & 0 &$ \frac{2 \gamma }{\sqrt{3}}$
   & 0 & $\frac{\sqrt{2} \gamma }{3}$ & 0 & $\frac{4
   \gamma }{3}$ \\
$D J/\psi$& -$\sqrt{2} \gamma$  & 0 & $\frac{2 \gamma }{\sqrt{3}}$ & 0 &
   $\frac{\sqrt{2} \gamma }{3}$ & 0 &$ \frac{4 \gamma }{3}$
   & 0
\end{tabular}

\hbox{    }

\hbox{    }

\center{\underline{S=-1, I$^G(J^{PC})$=$0(1^+)$}}

\begin{tabular}{c|cc}
Channels & $D\bar{K^*}$&$\bar{K}D^*$ \\
\hline
 $D\bar{K^*}$&-1 & -$\gamma$  \\
 $\bar{K}D^*$&-$\gamma$  &- 1
\end{tabular}

\hbox{    }

\hbox{    }

\subsection{Hidden-Charm Sector (C=0)}

\center{\underline{S=1, I$^G(J^{PC})$=$\oh(1^+)$}}

\begin{tabular}{c|cccccccc}
Channels & $\pi K^*$&$K\rho$&$\eta K^*$&$K\omega$&
 $\bar{D}D_s^*$&$D_s\bar{D^*}$&$KJ/\psi$&$\eta_c K^*$ \\
\hline
$\pi K^*$&-2 & $\frac{1}{2} $& 0 & $\frac{3}{2}$ & -$\sqrt{\frac{3}{2}}
   \gamma $ & 0 & 0 & 0 \\
 $K\rho$&$\frac{1}{2}$ & -2 & -$\frac{3}{2}$ & 0 & 0 &
  $ \sqrt{\frac{3}{2}} \gamma $ & 0 & 0 \\
$\eta K^*$& 0 &$ -\frac{3}{2}$ & 0 &$ \frac{3}{2} $& $-\frac{\gamma
   }{\sqrt{6}}$ & $\sqrt{\frac{2}{3}} \gamma $ & 0 & 0 \\
 $K\omega$&$\frac{3}{2}$ & 0 &$ \frac{3}{2}$ & 0 &
   $\sqrt{\frac{2}{3}} \gamma$  &$- \frac{\gamma }{\sqrt{6}}$
   & 0 & 0 \\
 $\bar{D}D_s^*$&$-\sqrt{\frac{3}{2}} \gamma $ & 0 &$ -\frac{\gamma
   }{\sqrt{6}}$ &$ \sqrt{\frac{2}{3}} \gamma $ &$ -\psi$ &
   0 &$- \frac{2 \gamma }{\sqrt{3}}$ & $-\frac{2 \gamma
   }{\sqrt{3}} $\\
 $D_s\bar{D^*}$&0 &$ \sqrt{\frac{3}{2}} \gamma $ & $\sqrt{\frac{2}{3}}
   \gamma$  & $-\frac{\gamma }{\sqrt{6}}$ & 0 & $-\psi $ &
   $-\frac{2 \gamma }{\sqrt{3}}$ & $-\frac{2 \gamma
   }{\sqrt{3}}$ \\
$KJ/\psi$& 0 & 0 & 0 & 0 & $-\frac{2 \gamma }{\sqrt{3}}$ & $-\frac{2
   \gamma }{\sqrt{3}}$ & 0 & 0 \\
$\eta_c K^*$& 0 & 0 & 0 & 0 & $-\frac{2 \gamma }{\sqrt{3}}$ & $-\frac{2
   \gamma }{\sqrt{3}}$ & 0 & 0
\end{tabular}

\hbox{    }

\hbox{    }

\center{\underline{S=0, I$^G(J^{PC})$=$1^+(1^{+-})$}}

\begin{tabular}{c|cccccc}
Channels & $\pi\omega$&$\eta\rho$&$\squd(\bar{K}K^*+c.c.)$&$\squd(\bar{D}D^*+c.c.)$&$\eta_c\rho$&$\pi J/\psi$ \\
\hline
$\pi\omega$& 0 & 0 & $\sqrt{3}$ &$ \frac{\gamma }{\sqrt{3}} $& 0 & 0 \\
 $\eta\rho$&0 & 0 & $\sqrt{3}$ & $\frac{\gamma }{\sqrt{3}}$ & 0 & 0 \\
$\squd(\bar{K}K^*+c.c.)$&$ \sqrt{3}$ & $\sqrt{3}$ & -1 & $-\gamma$  & 0 & 0 \\
$\squd(\bar{D}D^*+c.c.)$& $\frac{\gamma }{\sqrt{3}}$ &$ \frac{\gamma }{\sqrt{3}} $&
  $- \gamma$  & $-\psi$ & $2 \sqrt{\frac{2}{3}} \gamma  $& 2
  $ \sqrt{\frac{2}{3}} \gamma $ \\
 $\eta_c\rho$&0 & 0 & 0 & $2 \sqrt{\frac{2}{3}} \gamma$  & 0 & 0 \\
$\pi J/\psi$& 0 & 0 & 0 & $2 \sqrt{\frac{2}{3}} \gamma $ & 0 & 0
\end{tabular}

\hbox{    }

\hbox{    }

\center{\underline{S=0, I$^G(J^{PC})$=$1^-(1^{++})$}}

\begin{tabular}{c|cccccc}
Channels& $\squd(\bar{K}K^*-c.c.)$&$\pi\rho$&$\squd(\bar{D}D^*-c.c.)$ \\
\hline
 $\squd(\bar{K}K^*-c.c.)$&-1 &$ -\sqrt{2} $& $\gamma$  \\
 $\pi\rho$&$-\sqrt{2} $& -2 & $\sqrt{2} \gamma $ \\
 $\squd(\bar{D}D^*-c.c.)$&$\gamma$  & $\sqrt{2} \gamma$  & $-\psi$
\end{tabular}

\hbox{    }

\hbox{    }

\center{\underline{S=0, I$^G(J^{PC})$=$0^+(1^{++})$}}

\begin{tabular}{c|ccc}
Channels& $\squd(\bar{K}K^*+c.c.)$&$\squd(\bar{D}D^*+c.c.)$& $\squd(\bar{D_s}D_s^*-c.c.)$\\
\hline
 $\squd(\bar{K}K^*+c.c.)$& -3 & -$\gamma$  &$ \sqrt{2} \gamma$  \\
 $\squd(\bar{D}D^*+c.c.)$&$-\gamma $ & $-\psi -$2 & $-\sqrt{2} $\\
$\squd(\bar{D_s}D_s^*-c.c.)$&$\sqrt{2} \gamma $ &$ -\sqrt{2}$ &$ -\psi- $1
\end{tabular}

\hbox{    }

\hbox{    }

\hbox{    }

\hbox{    }

\center{\underline{S=0, I$^G(J^{PC})$=$0^-(1^{+-})$}}

\begin{tabular}{c|cccccccc}
Channels &$\pi\rho$&$\eta\omega$& $\squd(\bar{D}D^*$&$\eta_c\omega$&$\eta J/\psi$& $\squd(\bar{D_s}D_s^*$&$\squd(\bar{K}K^*$&$\eta_c J/\psi$ \\
& & &$-c.c.)$ & & & $+c.c.)$ &$-c.c.)$ &  \\
\hline
$\pi\rho$&-4 & 0 & $\sqrt{3} \gamma $ & 0 & 0 & 0 & $-\sqrt{3}$ & 0 \\
 $\eta\omega$&0 & 0 &$ -\frac{\gamma }{3}$ & 0 & 0 & $-\frac{2 \sqrt{2}
   \gamma }{3}$ & 3 & 0 \\
  $\squd(\bar{D}D^*-c.c.)$&$\sqrt{3} \gamma$  & $-\frac{\gamma }{3}$ & $-\psi-$2 &
  $ -\frac{2 \sqrt{2} \gamma }{3}$ & $-\frac{2 \sqrt{2} \gamma
   }{3}$ & $-\sqrt{2}$ & $\gamma $ & $-\frac{8 \gamma }{3}$ \\
 $\eta_c\omega$&0 & 0 &$- \frac{2 \sqrt{2} \gamma }{3} $& 0 & 0 &$ \frac{4
   \gamma }{3} $& 0 & 0 \\
 $\eta J/\psi$&0 & 0 &$ -\frac{2 \sqrt{2} \gamma }{3} $& 0 & 0 & $\frac{4
   \gamma }{3}$ & 0 & 0 \\
 $\squd(\bar{D_s}D_s^*+c.c.)$&0 &$ -\frac{2 \sqrt{2} \gamma }{3}$ & $-\sqrt{2}$ &$\frac{4
   \gamma }{3}$ &$\frac{4 \gamma }{3}$ & $-\psi -$1 &
  $\sqrt{2} \gamma $ &$- \frac{4 \sqrt{2} \gamma }{3}$ \\
 $\squd(\bar{K}K^*-c.c.)$&$-\sqrt{3}$ & 3 &$\gamma $ & 0 & 0 &$ \sqrt{2} \gamma  $& -3
   & 0 \\
$\eta_c J/\psi$& 0 & 0 & $-\frac{8 \gamma }{3}$ & 0 & 0 &$ -\frac{4 \sqrt{2}
   \gamma }{3} $& 0 & 0
\end{tabular}

\hbox{    }

\hbox{    }

\end{document}